\documentclass[a4paper,twocolumn,amsmath,amssymb,aps]{revtex4-2}
\usepackage[utf8]{inputenc}
\usepackage[T1]{fontenc}
\usepackage[english]{babel}
\usepackage{graphicx}
\usepackage{siunitx}
\usepackage{color}
\usepackage{hyperref}

\DeclareMathOperator{\e}{e}
\usepackage{xcolor}
\begin{document}

\title{Experimental test of scalar-field based dark energy models \\using neutron interferometry}
\author{Andreas Dvorak$^1$}
\email{andreas.dvorak@tuwien.ac.at}
\author{Kazuma Obigane$^{1,3}$}
\author{Hartmut Lemmel$^{1,2}$}
\author{Tobias Jenke$^2$}
\author{Stephan Sponar$^1$}
\email{stephan.sponar@tuwien.ac.at}
\affiliation{%
$^1$Atominstitut, TU Wien, Stadionallee 2, 1020 Vienna, Austria\\
$^2$Institut Laue Langevin, 38000 Grenoble, France\\
$^3$Division of Applied Physics, Graduate School of Engineering, Hokkaido University, Sapporo 060-8628, Japan}
\date{\today}

\begin{abstract}
We report phase shift measurements of neutron matter waves propagating in vacuum and low-pressure argon gas, using a technique developed for neutron interferometric scattering length measurements. The experiment probes additional phase shifts induced by couplings to scalar fields. Based on the absence of such effects, we derive stringent constraints on scalar symmetron and dilaton fields, both of which represent leading candidates for quintessence dark energy.
\end{abstract}

\maketitle

{\it Introduction.---}The source of dark energy, an unknown substance that manifests in the accelerated expansion rate of our Universe, is one of the most important unsolved questions in modern cosmology \cite{Riess:1998,Perlmutter:1999}. Within the standard $\Lambda$CDM cosmological model, dark energy is phenomenologically described by a cosmological constant, representing a spatially and temporally constant vacuum energy density responsible for the accelerated expansion of the Universe. A solution to this problem might be found in new hypothetical scalar fields with a self-interaction and a coupling to matter via a screening mechanism, accounting for dark energy \cite{Joyce:2014kja}. The scalar field acts effectively like a cosmological constant, thereby driving the observed accelerated expansion of the Universe. The corresponding effective potential is such that a fifth force induced by the scalar field would be strongly suppressed in regions of high mass densities like on Earth. On cosmological scales, however, the ambient mass densities are very low, and the force is no longer suppressed.

There are three prominent dark energy candidates with their scalar fields described by dilaton (D) \cite{Brax:2010gi}, symmetron (S) \cite{Cronenberg:2018qxf,Brax:2017hna,Pitschmann:2020ejb,Brax:2018grq} and chameleon models \cite{Burrage:2018}. The exponentially decreasing dilaton potential naturally arises in the strong-coupling limit of string theory \cite{Gasperini:2001pc,Damour:2002nv,Damour:2002mi} and has been proposed as a viable dark energy candidate. Initial experimental constraints have been obtained from Lunar Laser Ranging experiments \cite{Müller:2019}, gravity resonance spectroscopy \cite{Jenke2014,Jenke:2011zz}, atom-interferometry \cite{Hamilton:2015}, and projected constraints from Casimir-effect measurements  \cite{Sedmik:2021iaw}, as reported in Refs. \cite{Brax:2022uyh,Fischer:2023koa}. In addition, Ref. \cite{Kading:2023mdk} demonstrates that further bounds can be derived by analyzing dilaton-induced open quantum dynamics within the theoretical framework developed in Refs. \cite{Burrage:2018pyg,Burrage:2019szw,Kading:2022jjl}. A Symmetron model, on the other hand, relies on a mechanism of spontaneous symmetry breaking analogous to that of the Standard Model Higgs field \cite{Hinterbichler:2010es,Hinterbichler:2011ca}. 

These models predict a quantum-mechanical phase shift for neutrons traversing a vacuum region in which the corresponding scalar field develops in the absence of matter \cite{Brax:2013,Fischer:2024a}. The field depends not only on the remaining gas pressure but also on the distance to the chamber walls. Previous neutron interferometric measurements by the TU Wien/ILL group, are reported in \cite{Lemmel:2015kwa}. Results of the INDEX collaboration (NIST, NIOF, Indiana, Waterloo, NCSU) using dual aluminum chambers and mbar He pressure modulation, are reported in \cite{Snow:2015,Li:2016tux}, giving stronger chameleon bounds in the strongly coupled regime.
Regarding the chameleon fields, we do not expect to improve the stringent limits already known (see \cite{Burrage:2018} for a living review).

{\it Theory.---}The effective potentials of the scalar fields under consideration are given by
\begin{align}
    V_{\rm eff\,,D,S}(\phi,\rho) = V_{\rm D,S}(\phi)+\rho\,A_{\rm D,S}(\phi),
\end{align}
where $V(\phi)$ describes the self-interactions of the field, and the Weyl factor $A(\phi)$ couples to the ambient matter density $\rho$. The dilaton (D) and symmetron (S) models are defined in \cite{Brax:2018iyo} as
\begin{align}
    V_{\rm D}(\phi) &= V_0\e^{-\lambda_{\rm D}\phi/m_{\rm{pl}}},\\
    V_{\rm S}(\phi) &= -\frac{\mu^2}{2}\phi^2+\frac{\lambda_{\rm S}}{4}\phi^4,
\end{align}
together with
\begin{align}
    A_{\rm D}(\phi) &= 1+\frac{A_2}{2}\frac{\phi^2}{m_{\rm pl}^2},\\
    A_{\rm S}(\phi) &= 1+\frac{\phi^2}{2\,M^2},
\end{align}
resulting in the eﬀective potentials
\begin{align}
    V_{\rm{eff,D}}(\phi,\rho) &= V_0\e^{-\lambda_{\rm D}\,\phi/m_{\rm pl}}+A_2\,\rho\frac{\phi^2}{2\,m_{\rm pl}^2},\\
    V_{\rm eff,S}(\phi,\rho)  &= \frac{1}{2}\left(\frac{\rho}{M^2}-\mu^2\right)\phi^2+\frac{\lambda_{\rm S}}{4}\phi^4.
\end{align}
The dilaton parameters are the energy scale $V_0$, related to the dark energy of the Universe, the dimensionless parameter $\lambda_{\rm D}$ as well as the dimensionless coupling parameter $A_2$, while $m_{\rm pl}$ denotes the reduced Planck mass. The symmetron parameters are the dimensionless coupling parameter $\lambda_{\rm S}$ and the two mass scales $\mu$ and $M$. 
Following \cite{Fischer:2024a},
to neglect possible couplings to matter of higher order, we restrict our analysis to
\begin{align}
    \frac{A_2}{2}\frac{\phi^2}{m_{\rm pl}^2},\, \frac{\phi^2}{2\,M^2} \ll 1,
\end{align}
which leads to sharp cut-offs in the displayed exclusion plots. However, regions where neutron interferometry is sensitive and this condition does not hold have already been constrained by other experiments.

\begin{figure}[!t]
    \centering
    \includegraphics[width=0.48\textwidth]{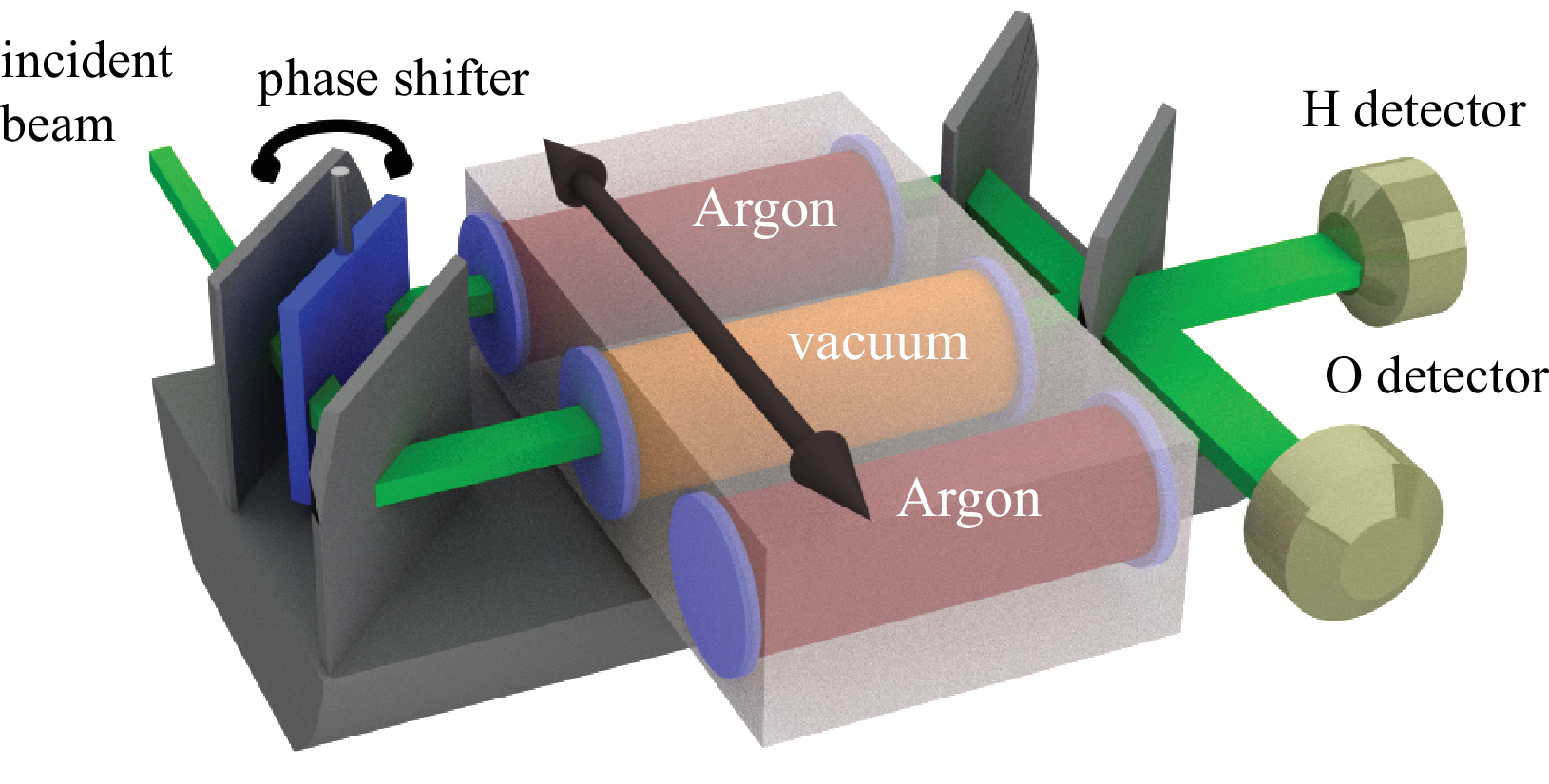}
    \caption{Schematic illustration of the neutron interferometric setup with phase shifter, argon chambers, vacuum chamber, and detectors.}
    \label{fig:schematic}
\end{figure}

\begin{figure}[!b]
    \centering
    \includegraphics[width=0.48\textwidth]{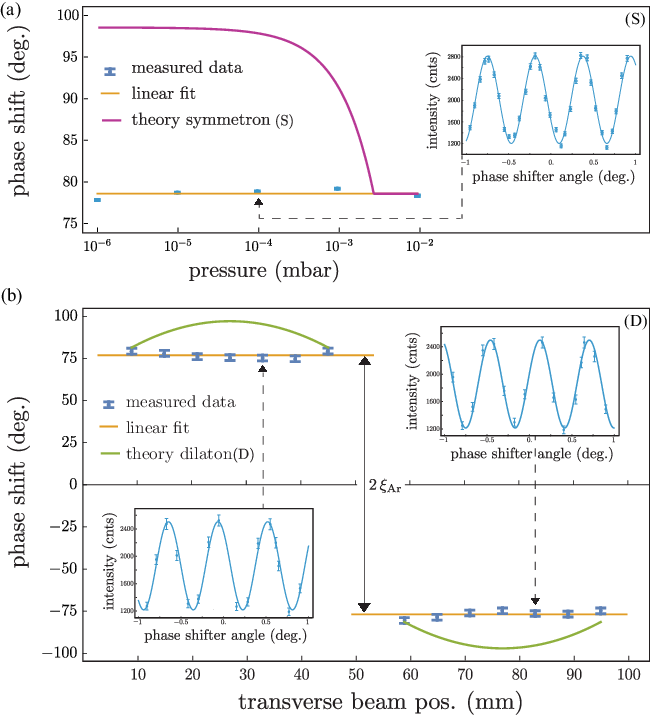}
    \caption{(a) Measured phase shifts vs. pressure in the vacuum chamber, together with phase predictions of the symmetron (S) model with parameters $\mu = \SI{0.1}{keV}$, $M = \num{4.47e-7}$, and $\lambda_{\rm S} = \num{4.76e7}$ (magenta line; see Fig.\,\ref{fig:results}(a) left panel for full exclusion plot). The Inset shows an interferogram at \SI{1.02e-4}{mbar}. (b) Measured phase shifts vs. transverse beam positions including phase predictions of the dilaton (D) model with parameters $V_0 = \SI{10}{MeV^4}$, $\lambda_{\rm D} = \num{e-5}$, and $A_2 = \sqrt{10} \times 10^{30}$ (green line), together with interferograms at two fixed transverse beam positions (insets). The exclusion plot for this model parameters can be seen in Fig.\,\ref{fig:results}\,(b). The distinct phase jump of $2\,\xi_{\rm Ar}$ close to position \SI{50}{mm} occurs when the vacuum chamber moves from one beam path to the other one.}
    \label{fig:phaseshifts}
\end{figure}

The experimental setup (Fig. \ref{fig:schematic}) is such that the neutron beam is split into two paths. Along each path, the neutrons traverse a gas chamber. One chamber is filled with argon gas at ambient pressure (argon chamber) while the other one is evacuated (vacuum chamber). This situation creates the argon phase
\begin{align}
    \xi_{\rm{Ar}}&=-N\,b_c\,\lambda\,D = \SI{-76\pm 2}{\degree} \label{chiAr}
\end{align}
where $N$ denotes the number density of argon atoms, $b_c$ the coherent neutron scattering length of argon atoms, $\lambda$ the neutron wave length and $D$ the chamber length, \cite{rauch:2015neutron}. This phase is well understood from neutron scattering length measurements \cite{Kaiser:1979,Haun:2020}.
The hypothetical scalar fields then create an additional phase which can be distinguished from the argon phase by qualitative features. First, while evacuating the vacuum chamber $\xi_{\rm Ar}$ is already established at about \SI{1e-2}{mbar} and stays constant when the vacuum is further improved. In contrast, the symmetron phase is expected to develop very abruptly below a certain threshold pressure due to a sudden phase transition. Second, the dilaton phase is expected to be strongest in the chamber center and suppressed close to the chamber walls. A transverse scan of the chamber would reveal such a pattern. Typical expected scalar field phases are shown in Fig.\,\ref{fig:phaseshifts}.

To leading order, the scalar-field–induced phase shift is given by \cite{Greenberger:1979zz}
\begin{align}
    \xi_{\rm D,S} = -\frac{m_n}{k_0}\int_{\rm CFP}U_{\rm D,S}\left(\mathbf{x}\right)\,{\rm d}s\,,
\end{align}
the integration extending over the classical flight path (CFP) and
\begin{align}
    U_{\rm D}(\mathbf{x}) &= \mathfrak{Q}_{\rm D}\frac{A_2\,m_n}{2\,m_{\rm pl}^2}\left(\phi^2(\mathbf{x})-\phi_{\rm gas}^2\right),\\
    U_{\rm S}(\mathbf{x}) &= \mathfrak{Q}_{\rm S}\frac{m_n}{2\,M^2}\left(\phi^2(\mathbf{x})-\phi_{\rm gas}^2\right).
\end{align}
Here, $k_0$ is the wave number of the neutron, $m_n$ its mass, $\phi_{\rm gas}$ is the field value that minimizes the potential in the argon gas and $\mathfrak{Q}_{\rm X}$ denotes the screening charge of the neutron. Computing the field of the neutron and the surrounding chamber generally constitutes a two-body problem involving the neutron and the vacuum or argon-filled cavity. We approximate this problem by modeling the neutron as a classical massive sphere with a radius of \SI{0.5}{fm}, consistent with QCD expectations. This approximation, referred to as Fermi screening, enables an effective treatment of neutron screening effects. Within this framework, the scalar potential is rescaled by a screening charge 
$\mathfrak{Q}$, which ranges from 0 for a fully screened sphere to 1 for an unscreened sphere. The analytical expressions can be found in \cite{Brax:2022uyh} for the dilaton and in \cite{Brax:2017hna,Pitschmann:2020ejb} for the symmetron model.

For each field model it is necessary to compute $\phi$ inside the vacuum and the argon chamber, which amounts to solving the non-linear differential equation (see \cite{Fischer:2024a})
\begin{align}
    \Delta\phi(\mathbf{x}) = V_{{\rm eff,}\phi}(\phi(\mathbf{x}),\rho(\mathbf{x}))\,.
\end{align}
Due to the cylindrical geometry of the vacuum chamber (see Fig.\,\ref{fig:schematic}), it follows that
\begin{align}
    V_{{\rm eff,}\phi}(\rho,\phi) = \left[\frac{1}{r}\frac{\partial}{\partial r}\left(r\frac{\partial}{\partial r}\right)+\frac{\partial^2}{\partial z^2}\right]\phi(r,z)\,.
\end{align}
As a boundary condition, we require that the scalar field $\phi$ reaches its potential minimum $\phi_{\rm M}$ within the shell of the chamber, and we restrict our analysis to parameter regimes in which this assumption is justified. Owing to the absence of analytical solutions, the corresponding differential equations are solved numerically.

\begin{figure*}
    \centering
    \includegraphics[width=0.85\textwidth]{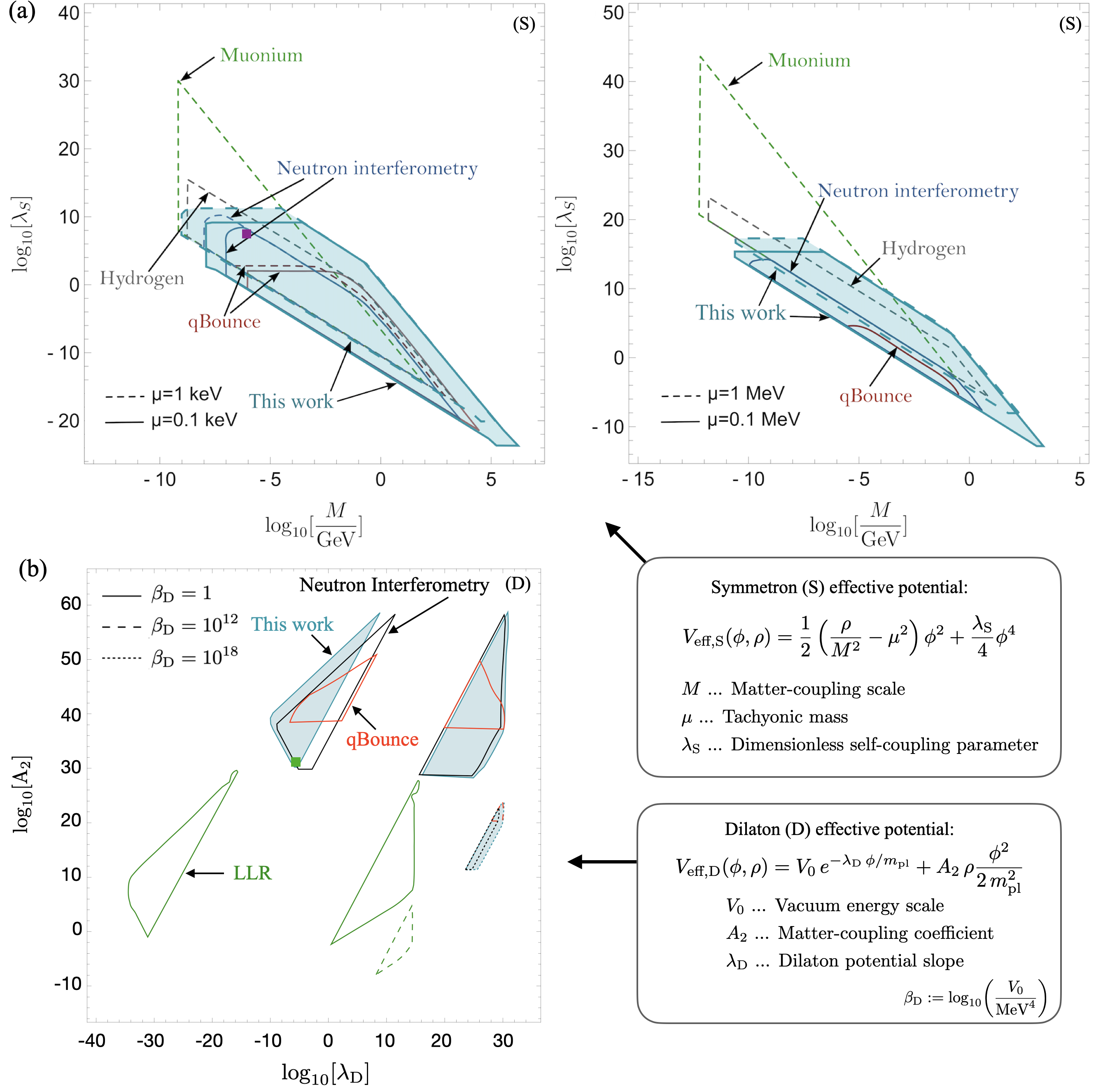}
    \caption{Exclusion plots for symmetrons (a) at different mass scales $\mu$ (left: keV, right: MeV). The present work (shaded in teal) is compared to Eöt-Wash results \cite{Upadhye:2013}, the atom interferometry analysis [{\citealp{Hamilton:2015}},\,\citealp{Panda:2024}], investigations of hydrogen, muonium and the electron \cite{Brax:2023}, analysis for qBounce, and previous neutron interferometry results \cite{Fischer:2024a,Fischer:2024b}. The magenta square marks model parameters used for the theory curve in Fig.\,\ref{fig:phaseshifts}\,(a). Exclusion plots for  dilatons (b) at different values of 
    $\beta_{\rm D}$. The present work (shaded in teal) is compared with previous interferometer, qBounce and Lunar Laser Ranging (LLR) results taken from \cite{Fischer:2023koa}. The green square marks model parameters used in Fig.\,\ref{fig:phaseshifts}\,(b).
    \label{fig:results}}
\end{figure*}

{\it Experiment.---}Unpolarized neutrons pass the two paths of a skew-symmetric Mach-Zehnder type interferometer, where the [220] lattice planes of silicon crystals are used in Laue geometry (lattice planes perpendicular to the surface of the crystal plates). The interferometer has a beam separation of \SI{50}{mm} over a length of \SI{160}{mm}, allowing to insert individual chambers in the two beam paths, as shown in Fig.\,\ref{fig:schematic}. The inner chamber length is \SI{100}{mm}.
A gas handling system sets a constant pressure of \SI{1040}{mbar} of pure argon gas in the two outer chambers. We chose argon rather than ambient air to ensure a well-defined gas composition.
The middle chamber is connected to a vacuum control system consisting of a pressure gauge, a motorized leak valve and two pumps. The pumps (pre-pump and turbomolecular pump) are operating continuously while a controlled amount of argon is admitted through the leak valve to maintain a vacuum down to approximately \SI{1e-6}{mbar}. By translating the chambers, the vacuum can be placed in the left or right beam path with one of the argon chambers as reference in the other path. Sapphire plates on the front and the back side serve as identical windows for all three chambers. Further experimental details are presented in the Supplemental Material \cite{Supp}.
The experiment was carried out at the neutron interferometer instrument S18 at the high-flux reactor of the Institute Laue-Langevin (ILL) in Grenoble, France. The crystal interferometer was operated at a Bragg angle of \SI{45}{\degree} resulting in a monochromatic beam with mean wavelength $\lambda=\SI{2.7}{\angstrom}$ $(\frac{\delta\lambda}{\lambda}\approx0.043)$. A $5\times5$~\SI{}{mm^2} beam cross section was used.

The interferograms were obtained by rotating an auxiliary phase shifter (see Fig.\,\ref{fig:schematic}) and recording the intensity oscillations between O and H detector. Such interferograms are measured before and after some parameter variation. The relative displacement of the corresponding sinusoidal curves represents the phase shift induced by the parameter change. The acquisition of each interferogram typically requires in the order of half an hour. During this time, the intrinsic phase of the interferometer can drift due to temperature fluctuations or other environmental influences. To compensate for such drifts, the motion of the phase shifter is interlaced with the variation of the experimental parameter. Specifically, the phase shifter is rotated to some angular position, and neutrons are counted for \SI{15}{s} at each transverse beam position. The phase shifter is then advanced to the next position, and neutron counting is repeated for all transverse beam positions, and so forth. In this manner, we obtain interferograms measured simultaneously for all parameter settings. Consequently, their relative phases are free of phase drifts.

We used the largest neutron interferometer available \cite{Zawinsky:2002} with a loop size of \SI{50}{mm} $\times$ \SI{160}{mm} in order to maximize the size of the chambers. Such large single crystal interferometers are extremely sensitive to temperature gradients, air flow, vibrations, bending, and other perturbations. Consequently, the interference contrast (fringe visibility) was limited to \SI{27}{\%} to \SI{45}{\%}. The interferograms appear somewhat noisier than what can be accounted for by pure counting statistics. This implies that the phase is slightly fluctuating over the duration of each interferogram acquisition. We conservatively accounted for this noise by performing a $\chi^2$ test for each sine fit and by inflating the fit error by a factor of about 2 such that the $\chi^2$ test was satisfied.

For each phase shifter position, we measured at seven transverse beam positions with the vacuum chamber in the left beam path and seven with the vacuum chamber in the right beam path. Each of these seven-point groups can be tested against bubble-like phase profiles predicted by dilaton models as illustrated in Fig. \ref{fig:phaseshifts} (b). In addition, the difference between left and right group gives twice the argon phase $\xi_{\rm Ar}$. This value, measured at different levels of vacuum, can be tested against predictions by dilaton models as illustrated in Fig. \ref{fig:phaseshifts} (a). Both tests use relative measurements which do not require high-precision determination of the argon phase itself. However, the measured value of $\xi_{\rm{Ar}}=\SI{77.67\pm0.14}{\degree}$ agrees with the theoretical one (Eq.\,\ref{chiAr}) which confirms the validity of the setup.

{\it Data acquisition and evaluation.---}
Based on the measurements acquired at 7 transverse beam positions per path a global multivariate fit was performed in which the phase was treated as the sole free parameter.
We calculated limits for $\lambda_{\rm D}$ and $\lambda_{\rm S}$ at fixed $A_2$ and $\beta_{\rm D} : =\log_{10}\!\left(V_0/{\rm MeV}^4\right)$, and $M$ and $\mu$, respectively, by comparing the calculated phase shifts $\xi$ with the measured phase shifts $\zeta\pm\sigma$.

We assume certain values of $\lambda_{\rm D}$ respectively $\lambda_{\rm S}$ and calculate the corresponding probability
\begin{align}
    p(\lambda_{\rm D,S})=\frac{\exp\left(-\frac{1}{2}\chi(\lambda_{\rm D,S})^2\right)}{\int_0^{\lambda_{\rm D,S}{_{\rm max}}}\exp\left(-\frac{1}{2}\chi(\lambda_{\rm D,S})^2\,{\rm d}\lambda_{\rm D,S}\right)},\\
    \chi(\lambda_{\rm D,S})^2=\sum_i\frac{\left(\xi(\lambda_{\rm D,S})_i-\zeta_i\right)^2}{\sigma_i^2}.
\end{align}
We determined the limit $\lambda_{\rm D,S}{_{\rm lim}}$ with \SI{95}{\%} confidence level by numerically solving the equation
\begin{align}
    \int_0^{\lambda_{\rm D,S}{_{\lim}}}p(\lambda_{\rm D,S})\,\rm d\lambda_{\rm D,S}=\SI{95}{\%}\,.
\end{align}
The calculation is repeated for different values of $A_2$ and $\beta_{\rm D}$, respectively

{\it Results.---}Exclusion plots for the symmetron model in the $M$-$\lambda_S$ plane are shown in Fig.\,3\,(a) for $\mu$ values in the keV and MeV range. Lower energy ranges (eV, meV) are presented in the Supplemental Material \cite{Supp}, Fig.\,S2.
For $\mu=\SI{1}{meV}$ (see Fig.\,S1\,a), as well as \SI{0.1}{eV} and \SI{1}{eV} (see Fig.\,S1\,b), our results confirm existing literature. For $\mu = \SI{0.1}{keV}$ and \SI{1}{keV} (see Fig.\,\ref{fig:results} a left), as well as \SI{0.1}{MeV} and \SI{1}{MeV} (see Fig.\,\ref{fig:results} a right), our results extend limits from the existing literature by about one order of magnitude.
Similarly, Fig.\,3\,(b) displays the excluded regions of the dilaton parameter space across different 
values of $\beta _{\rm D}$. The obtained results extend limits from the existing literature for $\beta _{\rm D}=1$ and $\beta _{\rm D}=10^{18}$ clearly.
Both plots compare our results with previous experimental bounds as described in detail in the figure caption.
The teal-shaded regions obtained in this work demonstrate a significant improvement, both for symmetrons and dilatons.

{\it Conclusions.---}We see no additional phase shift compared to that of argon. We use this to place bounds on possible values for symmetron parameters $M$, $\lambda_{\rm S}$ and $\mu$, as well as dilaton parameters $\lambda_{\rm D}$, $A_2$ and $\beta_{\rm D}$, and created exclusion bands as shown in Fig.\,\ref{fig:results}. The novelty of this work lies in the systematic investigation of previously unexplored parameter regimes. Our results provide new physical insights and may serve as a reference for future theoretical and phenomenological studies.

{\it Acknowledgments.---}This research was funded in part by the Austrian Science Fund FWF (grant DOIs: 10.55776/P34239 and 10.55776/PAT3585625), the Austrian Research Promotion Agency FFG (grant  FO999896034) and  and JST SPRING (grant Number: JPMJSP2119). The authors acknowledge the ongoing hospitality of the ILL, Grenoble (France). The authors would like to thank the mechanical workshop at Atominstitut, TU Wien for their expert craftsmanship and precision in manufacturing the vacuum chamber. We are also grateful to S. Schneider for contributing both essential components and valuable expertise to the vacuum system. For open access purposes, we have applied a CC BY public copyright license to any authors' accepted manuscript version arising from this submission. The authors acknowledge TU Wien Bibliothek for financial support through its Open Access Funding Programme.

{\it Data availability.---}The data that support the findings of this study are openly available \cite{ILL-DATA.3-16-18}.

%

\clearpage
\onecolumngrid

\setcounter{equation}{0}
\setcounter{figure}{0}
\setcounter{table}{0}
\setcounter{section}{0}
\setcounter{page}{1}
\makeatletter
\renewcommand{\theequation}{S.\arabic{equation}}
\renewcommand{\thefigure}{S.\arabic{figure}}
\renewcommand{\thetable}{S.\arabic{table}}
\renewcommand{\thesection}{S.\arabic{section}}
\renewcommand{\arraystretch}{1.5}
\makeatother

\begin{center}
    \textbf{\large Supplemental Material: Experimental test of scalar-field based dark energy models using neutron interferometry}\\[1em]
    Andreas Dvorak$^{1,*}$, Kazuma Obigane$^{1,3}$, Hartmut Lemmel$^{1,2}$, Tobias Jenke$^{2}$, and Stephan Sponar$^{1,\dagger}$\\[0.5em]
    \small\textit{$^1$Atominstitut, TU Wien, Stadionallee 2, 1020 Vienna, Austria\\
    $^2$Institut Laue Langevin, 38000 Grenoble, France\\
    $^3$Division of Applied Physics, Graduate School of Engineering, Hokkaido University, Sapporo 060-8628, Japan}
\end{center}
\vspace{1em}

\begin{quote}
\small
This supplemental material gives a more detailed description of the experimental setup and the systematic error mitigation for the neutron interferometry experiment.
Furthermore, we present for the symmetron model additional results in the meV and eV range. 
\end{quote}
\vspace{1em}

\section{Details on the experimental setup}
\begin{figure*}[!h]
    \centering
    \includegraphics[width=0.85\textwidth]{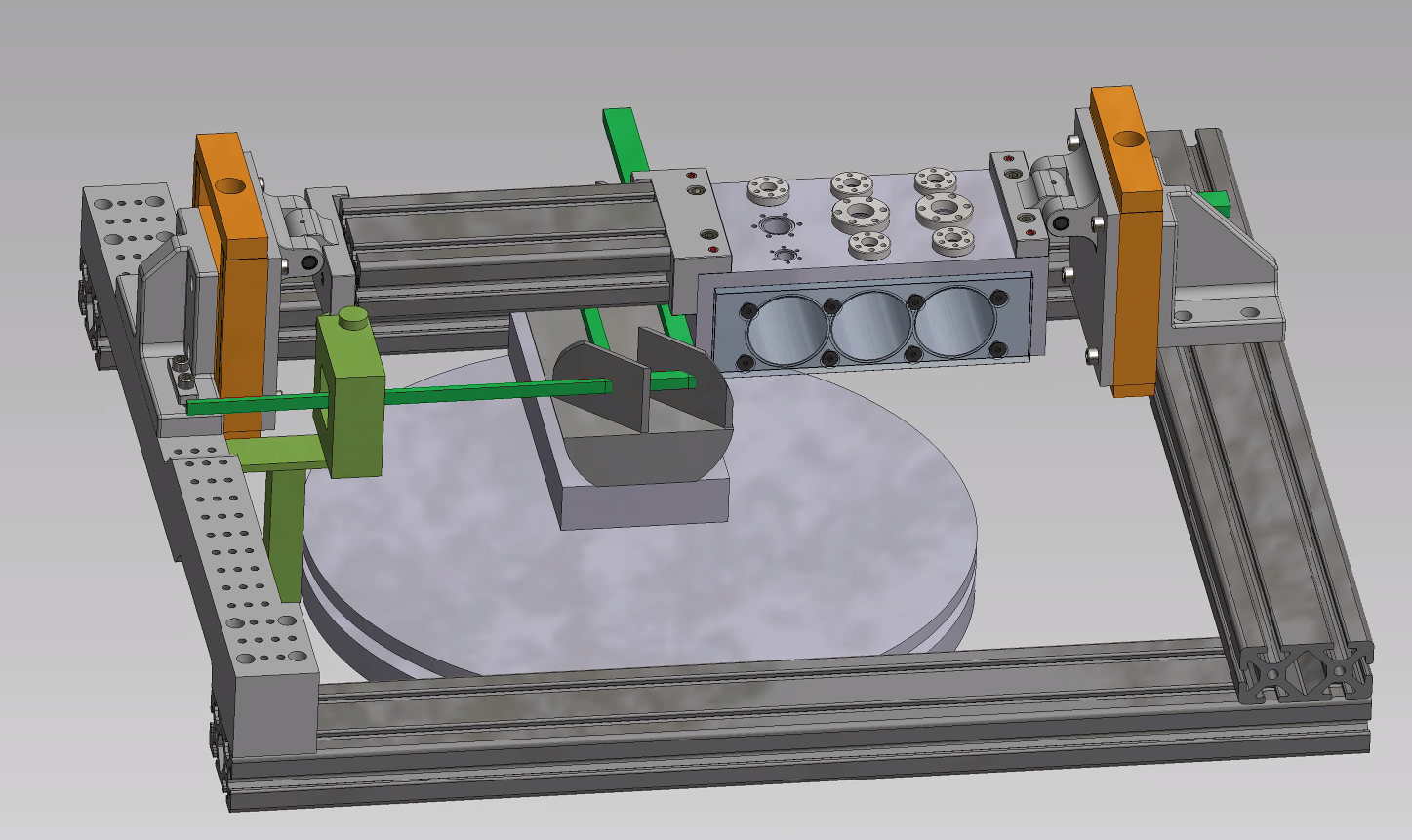}
    \caption{Technical drawing of the neutron interferometric
setup with silicon-interferometer, argon chambers (outer chambers), vacuum chamber (middle chamber), and the neutron beam (green).}
    \label{fig:TechnicalDrawing}
\end{figure*}

Unpolarized neutrons pass the two paths of a skew-symmetric Mach-Zehnder type interferometer, where the [220] lattice planes of silicon crystals are used in Laue geometry (lattice planes perpendicular to the surface of the crystal plates). The interferometer has a beam separation of \SI{50}{mm} over a length of \SI{160}{mm}, allowing to insert individual chambers in the two beam paths, as shown in Fig.\,\ref{fig:TechnicalDrawing}. The chambers have an inner diameter of \SI{47}{mm} and a length of \SI{100}{mm}.
A gas handling system (not depicted) sets a constant pressure of \SI{1040}{mbar} of pure argon gas in the two outer chambers. We chose argon rather than ambient air to ensure a well-defined gas composition.
The middle chamber is connected to a vacuum control system consisting of a pressure gauge, a motorized leak valve and two pumps. The pumps (pre-pump and turbomolecular pump) are operating continuously while a controlled amount of argon is admitted through the leak valve to adjust the vacuum pressure within the range of \SI{1E-6}{} to \SI{1E-2}{mbar}. By translating the chambers, the vacuum can be placed in the left or right beam path with one of the argon chambers as reference in the other path. \textcolor{black}{Special attention is payed to the mechanical stability of the translation stage. It has to account for the mechanical forces transmitted by the  bellow tube between the chambers and the turbomolecular pump. These forces vary as function of chamber position and pressure. The angular stability of the chamber during the movement is commissioned by laser reflection. The gap between the chamber windows and the interferometer lamellae is about two mm, likewise the gap between the bottom of the chamber and the interferometer base. 
Futher attention is payed to the thermal stability, since varying heat load onto the interferometer crystal, in particular as function of the chamber position, could create systematic errors. The turbomolecular pump is placed about one meter above the setup. A thermal shield between pump and setup ensures that no heat radiation of the pump reaches the interferometer. A combined Pirani and cold-cathode pressure gauge is mounted directly on the chamber. Its creates a small heat load onto the chamber which is cooled away by water tubes.}
The chambers are closed with \SI{7}{mm} thick sapphire plates on both sides ensuring identical windows for all three chambers. The plates have a thickness gradient of about \SI{4.4}{\micro\meter} over the whole length which creates a systematic phase change of \SI{46.7\pm0.4}{\degree}, over the 14 measured chamber positions. A calibration scan recorded with all three chambers at equal pressure is subtracted from all other measurements. When a cell is evacuated the pressure on the windows creates a calculated bending of about \SI{40}{nm} in the window center. The resulting shortening of the chamber by \SI{80}{nm} in total is negligible compared to the chamber length of \SI{100}{mm}.

The experiment was carried out at the neutron interferometer instrument S18 at the high-flux reactor of the Institute Laue-Langevin (ILL) in Grenoble, France. The perfect
crystal silicon interferometer was operated at a Bragg angle of \SI{45}{\degree} and a monochromatic beam with mean wavelength $\lambda=\SI{2.7}{\angstrom}$ $(\frac{\delta\lambda}{\lambda}\approx0.043)$ and $5\times5$~\SI{}{mm^2} beam cross section.

\section{Additional symmetron results}

Exclusion plots for the symmetron model in the $M$-$\lambda_S$ plane are shown in Fig.\,\ref{fig:results} for $\mu$ values in the meV and eV range, complementing the results for the keV and MeV range presented in the main text. The obtained exclusion regions confirm the ones from existing literature, as described in the figre caption. 

\begin{figure*}[!h]
    \centering
    \includegraphics[width=0.96\textwidth]{FigS2.png}
    \caption{Exclusion plots for symmetrons at different mass scales $\mu$. \textcolor{black}{The present work (shaded in teal) is compared to} Eöt-Wash results \cite{Upadhye:2013}, the atom interferometry analysis [\textcolor{black}{\citealp{Hamilton:2015}},\,\citealp{Panda:2024}], investigations of hydrogen, muonium and the electron (g-2) \cite{Brax:2023}, analysis for qBounce, and \textcolor{black}{previous} neutron interferometry results \cite{Fischer:2024a,Fischer:2024b}.  }
    \label{fig:results}
\end{figure*}

\end{document}